\documentclass{article}
\usepackage{graphicx}
\usepackage{amsmath}

\input{tcilatex}

\begin{document}

\title{Determinantal invariant gravity}


\author{Nurettin Pirin{\c c}{\c c}io{\~g}lu}

\maketitle

\begin{abstract}
Einstein-Hilbert action with a determinantal invariant has been
considered. The obtained field equation contains the \texttt{inverse
Ricci tensor}, $\Re_{\alpha\beta}$. The linearized solution of
invariant has been examined, and constant curvature space-time
metric solution of the field equation gives different curvature
constant for each values of $\sigma$. $\sigma=0$ gives a trivial
solution for constant curvature, $R_{0}$.



\end{abstract}

\section{Introduction}\label{intro}

Observations of the universe accumulate many investigation on
Einstein theory of general relativity. One of them is the
modification of Einstein-Hilbert (EH) action. There are numerus
investigations on the modified EH action with different context
\cite{review1}. Such modifications cast a vital role in the
inflationary cosmological model of Starobinsky
\cite{starobinsky1980}. Modeling the exponentially expanding the
early universe, \textit{i.e.} the inflation, is the most capable
theory to explain the natural structure of the present universe;
such as horizon, flatness, isotropy, homogeneity \textit{etc.}.
There are various inflationary models of the universe were
introduced by different studies with different context
\cite{{starobinsky1979},{guth1981},{linde1984},{steinhardt},{shaposhnikov2007}}.
For more information one can see the review \cite{review2} and
references there in. According to the Planck observations
\cite{planck2013}, the most working model of inflation is that of
Starobinsky. Our goal in this paper is to construct a determinantal
invariant parameter which can be used in the EH action. The constant
curvature solution of the invariant coincides with Starobinsky
cosmological inflationary model. Also, constant curvature space-time
solution of equation has been examined.

One can construct such a determinantal invariant with the same
analogy in \cite{{durmush2009},{pir1}}. The ratio of determinant of
the Ricci tensor and metric tensor \cite{weinberg} is

\begin{equation}
\hbox{r}=\frac{\widetilde{R}}{g} \label{det}
\end{equation}

Where $\widetilde{R}$ is the determinant of Ricci tensor,
$R_{\mu\nu}$, $g$ is that of metric tensor, $g_{\mu\nu}$. This
parameter is our determinantal invariant. Accordingly a parameter
function can be given as

\begin{equation}
\hbox{f(r)}=\xi\frac{(\hbox{r})^{\sigma}}{M^{4(2\sigma-1)}} \label{f}
\end{equation}

Where $\sigma$, $\xi$ are dimensionless constant numbers. $\sigma$
is relating mass parameter, $M$, and determinantal parameter,
$\widetilde{R}$ to fix $\hbox{f}$ as a dimensionfull parameter. One
can construct an action integral of EH with the determinantal
invariant function, $\hbox{f(r)}$, as follows

\begin{equation}
S=\int \hbox{d}^{4}\hbox{x} \sqrt{-g}\{\frac{M^{2}_{pl}}{2}R
+\xi\frac{1}{M^{4(2\sigma-1)}}(\frac{\widetilde{R}}{g})^{\sigma}\}
+S_{matter}. \label{action}
\end{equation}

Where $R$ is curvature scalar. Variation of equation
(\ref{action}) with respect to the metric tensor produces the
field equation

\begin{eqnarray}
G_{\mu\nu}+\frac{1}{M^{2}_{pl}}g_{\mu\nu}(2\hbox{rd}-1)\hbox{f(r)}+
\frac{\sigma}{M^{2}_{pl}}\{g_{\mu\nu} \nabla_{\alpha} \nabla_{\beta}[\hbox{f}\Re^{\alpha\beta}]+
\nabla_{\alpha} \nabla^{\alpha}[\hbox{f}\Re_{\mu\nu}]-\nonumber\\
2\nabla_{\mu} \nabla_{\alpha}[\hbox{f}\Re_{\nu}^{\alpha}]\}=\frac{1}{M^{2}_{pl}}T_{\mu\nu}.\label{field}
\end{eqnarray}

Where $\hbox{d}$ is derivative with respect to the determinantal
invariant, $\hbox{r}$, and
$G_{\mu\nu}=R_{\mu\nu}-\frac{1}{2}g_{\mu\nu}R$ is the Einstein
tensor. The equation (\ref{field}) has the novel structure, because
it contains \textit{inverse Ricci} tensor, $\Re^{\alpha\beta}$.
Comparing with the $f(R)$ gravity theories \cite{review1}, our field
equation is very different, because of it has three extra terms with
\textit{inverse Ricci} tensor, $\Re^{\alpha\beta}$. $\sigma=0$ case,
simplifies the field equation as follows

\begin{equation}
G_{\mu\nu}-\xi\frac{M^{4}}{M^{2}_{pl}}g_{\mu\nu}=\frac{1}{M^{2}_{pl}}T_{\mu\nu}.
\end{equation}

The vacuum solution of this equation is the maximally symmetric
solution of field equation (\ref{field}).

\section{Linearized solution}\label{sec:1}

In the linearized approximation (up to the first order of
$h_{\mu\nu}$) the metric tensor $g_{\mu\nu}$, and its inverse
$g^{\mu\nu}$ become

\begin{equation}
g_{\mu\nu}=\eta_{\mu\nu}+h_{\mu\nu}, \label{lin1}
\end{equation}

\begin{equation}
g^{\mu\nu}=\eta^{\mu\nu}-h^{\mu\nu}+\frac{1}{2}h^{\mu\alpha}h_{\alpha}^{\nu}.
\end{equation}

Where $\eta_{\mu\nu}=\hbox{diag}(-1,1,1,1)$ is the flat Minkowski
space-time metric. In the Minkowski background the linearized
Ricci tensor, and curvature scalar, become
\begin{equation}
R_{\mu\nu}=\frac{1}{2}(\partial_{\alpha}\partial_{\mu}h^{\alpha}_{\nu}+
\partial_{\alpha}\partial_{\nu}h^{\alpha}_{\mu}-\partial_{\mu}\partial_{\nu}h-
\partial^{\alpha}\partial_{\alpha} h)
\end{equation}
\begin{equation}
R=g^{\mu\nu}R_{\mu\nu}=\partial_{\mu}\partial_{\nu}h^{\mu\nu}-
\partial^{\alpha}\partial_{\alpha} h
\end{equation}

respectively. In this section we consider the behavior of field
equation (\ref{field}) in the linearized approximation. The
linearized form (expanding $\hbox{f}$ up to the firs order of
$h_{\mu\nu}$) of determinantal invariant is

\begin{equation}
\hbox{f}=-\frac{\widetilde{R}_{lin}}{M^{4}(1+h+...)}\approx -M^{-4}\widetilde{R}_{lin}=\hbox{f}_{lin} \label{det1}
\end{equation}

for $\sigma=1$. Where $\widetilde{R}_{lin}$ is the determinant of
the linearized Ricci tensor. Using empty space condition for
energy momentum tensor of matter, $T_{\mu\nu}=0$,  the linearized
solution \cite{{pir1},{zurab2007}} of equation (\ref{field}) in
the Minkowski background, expanding determinantal potential about
$\hbox{r}=0$, is obtained as

\begin{equation}
G_{\mu\nu}^{lin}= \frac{1}{M^{2}_{pl}}t_{\mu\nu} \label{field1}
\end{equation}

Where $G_{\mu\nu}^{lin}$ is the linearized Einstein tensor, and
$t_{\mu\nu}$ is the 1st order perturbed (gravitational field)
energy momentum tensor.

\begin{eqnarray}
G_{\mu\nu}^{lin}=\frac{1}{2}(\partial_{\alpha}\partial_{\nu}h^{\alpha}_{\mu}+
\partial_{\alpha}\partial_{\mu}h^{\alpha}_{\nu}-\partial_{\mu}\partial_{\nu}h-\nonumber\\
\partial^{\alpha}\partial_{\alpha}h_{\mu\nu}-
\eta_{\mu\nu}\partial_{\alpha}\partial_{\beta}h^{\alpha\beta}+
\eta_{\mu\nu}\partial^{\alpha}\partial_{\alpha}h)
\end{eqnarray}

The linearized Einstein field (\ref{field1}) equation takes the
form of

\begin{eqnarray}
\partial_{\alpha}\partial_{\nu}h^{\alpha}_{\mu}+
\partial_{\alpha}\partial_{\mu}h^{\alpha}_{\nu}-\partial_{\mu}\partial_{\nu}h-\nonumber\\
\partial^{\alpha}\partial_{\alpha}h_{\mu\nu}-
\eta_{\mu\nu}\partial_{\alpha}\partial_{\beta}h^{\alpha\beta}+
\eta_{\mu\nu}\partial^{\alpha}\partial_{\alpha}h=\frac{2}{M^{2}_{pl}} t_{\mu\nu} \label{graviton}
\end{eqnarray}

\section{Constant curvature space-time solution}\label{sec:2}

The space-time metric with constant curvature, $R_{0}$, is
characterized by the condition

\begin{equation}
R_{\mu\nu\alpha\beta}=\frac{1}{12}R_{0}(g_{\mu\alpha}g_{\nu\beta}-g_{\mu\beta}g_{\nu\alpha})
\label{reimann}
\end{equation}

on the Reimann tensor. So, the Ricci tensor satisfies

\begin{equation}
R_{\mu\nu}=\frac{1}{4}R_{0}g_{\mu\nu}. \label{ricci}
\end{equation}

from this, one can readily find the \textit{inverse Ricci} tensor as
follows

\begin{equation}
\Re^{\mu\nu}=\frac{4}{R_{0}}g^{\mu\nu}. \label{inversericci}
\end{equation}

The maximally symmetric solution of equation (\ref{field}) in vacuum
is

\begin{equation}
\frac{1}{4}R_{0}g_{\mu\nu}-\frac{2}{M^{2}_{pl}}g_{\mu\nu}r\hbox{f}_{r}+\frac{1}{M^{2}_{pl}}g_{\mu\nu}\hbox{f}=0.
\label{field2}
\end{equation}

Where

\begin{equation}
\hbox{r}=\frac{1}{256}R^{4}_{0},
\end{equation}

\begin{equation}
\hbox{f(r)}=\xi(\frac{R_{0}}{4M^{2}})^{4\sigma}M^{4},
\end{equation}

and

\begin{equation}
d\hbox{f}=\hbox{f}_{r}=\sigma\xi(\frac{R_{0}}{4M^{2}})^{4(\sigma-1)}\frac{1}{M^{4}}.
\end{equation}

Contracting the equation (\ref{field2}), gives us the algebraic

\begin{equation}
R_{0}-\xi\frac{4}{M^{2}_{pl}}(\frac{R_{0}}{4M^{2}})^{4\sigma}M^{4}(2\sigma-1)=0
\label{algebraic}
\end{equation}

equation. The solution for $R_{0}$ is not trivial for all values of
$\hbox{f}\neq 0$. But, one can get the trivial value of $R_{0}$ for
$\sigma=1/2$. $\sigma=0$ gives us the coupling constant $\xi$ which
linearly related to the $R_{0}$ as follows

\begin{equation}
\xi=-\frac{M^{2}_{pl}}{4M^{4}}R_{0}.
\end{equation}

This coupling constant is positive just for negative constant
curvature, $R_{0}$. Setting $\sigma=1$, $\xi$ becomes function of
constant curvature, and Planck mass

\begin{equation}
\xi=M^{2}_{pl}M^{4}(\frac{4}{R_{0}})^{3}.
\end{equation}

This is positive just for positive values of $R_{0}$.

In the case of $\sigma=1/2$, the determinantal invariant,
$\hbox{f}$, for 4-dimensional constant curvature space-time becomes

\begin{equation}
\hbox{f}_{c}=\xi\frac{1}{16}R_{0}^{2}. \label{det2}
\end{equation}

This result is compatible with Starobinsky inflationary model
\cite{starobinsky1980}, $R^{2}$. Then the constant curvature
solution of equation (\ref{field}) is

\begin{equation}
-M_{pl}^{2}g_{\mu\nu}R_{0}-\xi\frac{1}{8}g_{\mu\nu}R_{0}^{2}=0.\label{cc}
\end{equation}

One can compare this result with the special case (constant
curvature space-time) of Starobinsky inflationary parameter. The
Starobinsky model of inflation can be written as follows

\begin{equation}
\hbox{f}_{s}(R)=R-\frac{1}{6m^{2}}R^{2}. \label{starobinsky}
\end{equation}

This is known as the chaotic inflationary model of Starobinsky, and
it is perfectly well fitted with Planck data \cite{planck2013}. From
the comparison of inflationary parameters of equation (\ref{cc})
with that of Starobinsky, the inflation mass can be given as

\begin{equation}
m\simeq M_{pl}/\sqrt{\xi}. \label{mass}
\end{equation}

Inflaton mass \cite{brout} can be related to the reduced Planck mass
with $\xi\sim 1$ limit in the early universe.

\section{Conclusion}

Determinantal invariant modification of EH action (\ref{det}), does
not affect the linearized solution, equation (\ref{graviton}).
However, constant curvature space-time solution of EH action with
determinantal invariant, equation (\ref{det2}), mimics the
Starobinsky inflationary parameter, $R^{2}$, equation
(\ref{starobinsky}). The maximally symmetric solution of action
(\ref{action}) gives us very different results for coupling
constant, $\xi$.  Field equation (\ref{field}) contains inverse
Ricci tensor. Thus, the field equation (\ref{field}) may produce
novel results for physical or mathematical problems considered. As a
result one can guess the mass of inflaton from equation
(\ref{mass}).

\end{document}